\documentclass[reprint,aps,prl,floats]{revtex4-1}

\usepackage[dvips]{color}
\usepackage{graphicx}
\usepackage{bm}
\usepackage{textcomp} 

\begin{document}

\title{Fundamental limits on the losses of phase and amplitude optical actuators}

\author{Simone Zanotto}
\email{simone.zanotto@nano.cnr.it}
\affiliation{Dipartimento di Elettronica, Informazione e Bioingengeria, Politecnico di Milano, P.za Leonardo da Vinci 32, 20133 Milano, Italy.}
\altaffiliation{Present address: Istituto Nanoscienze - CNR, and Laboratorio NEST, Piazza San Silvestro 12, 56127 Pisa, Italy}

\author{Francesco Morichetti}
\affiliation{Dipartimento di Elettronica, Informazione e Bioingengeria, Politecnico di Milano, P.za Leonardo da Vinci 32, 20133 Milano, Italy}

\author{Andrea Melloni}
\affiliation{Dipartimento di Elettronica, Informazione e Bioingengeria, Politecnico di Milano, P.za Leonardo da Vinci 32, 20133 Milano, Italy}

\date{\today}

\begin{abstract}
Amplitude and phase are the basic properties of every wave phenomena; as long as optical waves are concerned, the ability to act on these variables is at the root of a wealth of switching devices. To quantify the performance of an optical switching device, an essential aspect is to determine the tradeoff between the insertion loss and the amplitude or phase modulation depth. Here it is shown that every switching optical device is subject to such a tradeoff, intrinsically connected to the dielectric response of the materials employed inside the switching element itself. This limit finds its roots in fundamental physics, as it directly derives from Maxwell's equations for linear dielectrics, and is hence applicable to a wide class of optical components. Furthermore it results that concepts as filtering, resonance and critical coupling could be of advantage in approaching the limit.
\end{abstract}

\maketitle

\section{Introduction}
In general, the performance of a device is intimately connected through fundamental physical laws to the properties of the materials or the sub-elements employed in its realization, and these connections may have far reaching implications to whole branches of engineering. For instance, the energy conversion efficiency of a solar cell is limited by several fundamental limits. For what concerns the photoexcited carrier exploitation, the Shockley-Queisser limit applies \cite{ShockleyJAP1961}; the photon trapping inside the absorber is instead ruled, in the ray-optics regime, by the Yablonovitch limit \cite{Yablonovitch} or by more general formulas recently proposed by Fan \textit{et al.} \cite{FanPRL2012} for wavelength-size patterned cells. 

Focusing back on the optical science, and more specifically on the integrated optical devices framework, recent developments are moving towards reconfigurable systems constituted by several elements, in order to implement complex operations on classical or quantum signals \cite{MillerPhotRes2013, PeruzzoNatComm2014}. As basic building blocks operating on the amplitude or on the phase of the wave, besides traditional switching elements -- like those relying on thermic, electric, or plasma dispersion effects -- devices involving novel materials are under investigation in the present years. Among them, it can be cited $\mathrm{VO}_2$ \cite{PoonAPL2013, PoonOE2012, AtwaterOE2010, WeissOE2012, OoiNanophotonics2013, JoushaghaniOE2015}, GST ($\mathrm{Ge}_2 \mathrm{Sb}_2 \mathrm{Te}_5$) \cite{TsudaOE2012, PruneriAPL2013, PerniceAPL2012, PerniceAdvMat2013, RudeACSPhot2015, AKUMNanoLett2013}, ITO (Indium Tin Oxide) \cite{Volker, AtwaterNanoLett2010}, polymeric materials \cite{LeutholdNatPhot2014}, and resistive switches \cite{LeutholdOptica2014}. With these materials, and in connection to other concepts like plasmonic waveguides, it is expected that certain device metrics like miniaturization, speed, energy consumption, and state retention, will be improved\cite{Volker_Review}. However, advantages usually come at a price, and in the present case this can be globally summarized as large losses.  
\begin{figure*}[ht!]
\centering
\includegraphics[width=\linewidth]{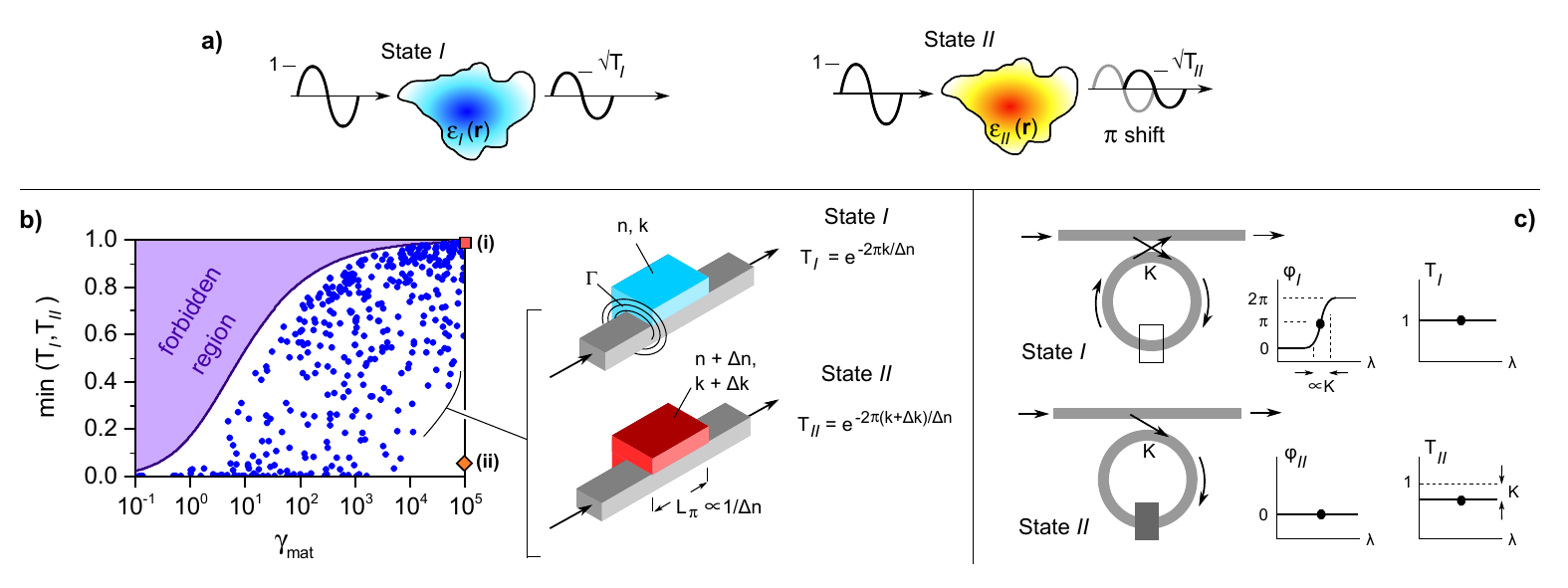}
\caption{Fundamental limit for a phase-switching optical element. (a) Schematic of the switching action. (b) Minimum transmission for a $\pi$-switch as a function of the material figure of merit $\gamma_{mat}$. The points describe the action of a simple device consisting of a waveguide loaded by the active material; different points correspond to different parameters $n$, $\Delta n$, $\kappa$ and $\Delta \kappa$. All the devices lie in the allowed region of the chart; however, certain devices are strongly sub-optimal. (c) Possible implementation of an optimized \textit{phase} switch based on a material which has \textit{intensity} switching properties.}
\end{figure*}

For instance, plasmonic waveguides systematically suffer from large losses, especially in the visible- and near-infrared spectral range, which is of interest for communications \cite{BoltassevaJOSAB2015}. This does not occur by chance, since the field confinement and the propagation losses are connected by a fundamental relation involving the sole properties of the plasmonic material, and hence of the noble metal optical constants \cite{ArbabiArxiv2014}. The presence of fundamental limits in optics, however, does not only concern guiding elements: considering intensity modulators, it has been recently highlighted that, when graphene is the active material, the insertion loss of the overall device is substantially governed by the graphene conductivity tensor, according to an inequality proved for planar, multilayered devices embedding conducting sheets \cite{TamagnoneNatPhot2014}. In this article we generalize that result, proving the existence of a lower limit also on the insertion losses introduced by a phase actuator. Moreover, our result applies in general to every two-port device with arbitrary geometry, like realistic structures in integrated optics. Data reported in the literature are critically analyzed in view of the present theory, and the role of resonance in switching devices is highlighted. A material figure of merit, depending on the sole dielectric constants of the switching material, turns out to be the central quantity for both amplitude and phase switches.

\section{Fundamental limit on the losses of phase actuators}

The first problem we address is to determine a fundamental limit on the insertion loss of a phase modulator. In its simplest implementation, its schematic is given in Fig.~1 (a). 
It is a two-port linear device that, when passing from state $I$ to state $II$, switches the phase of the output beam by a certain amount. Here we focus on the case of a $\pi$ switch, which is of relevance in most applications. While an ideal phase switch does not act on the amplitude, a real device possibly does that. Such a loss may be due to back-reflection, to scattering into other channels, or to absorption inside the switching region. Following the theory outlined in \cite{TamagnoneNatPhot2014, SchaugPettersenIRE1959}, and proved in the Supplementary Material for a device of arbitrary geometry,  it turns out that the insertion losses are ultimately determined by the sole complex permittivity of the switching material employed in the device. In formulas, 
\begin{equation}
\frac{4\ \mathrm{min}[T_I,T_{II}]}{(1-\mathrm{min}[T_I,T_{II}])^2} \le  \smash{\displaystyle\max_{\mathbf r \in V} \frac{|\varepsilon_I (\mathbf{r}) - \varepsilon_{II} (\mathbf{r})|^2}{4\  \varepsilon_I'' (\mathbf{r}) \ \varepsilon_{II}'' (\mathbf{r})} } \equiv \gamma_{mat}
\label{limit_pi_modulator}
\end{equation}
where $T_{I,II}$ are the intensity transmittance of the device in states $I$ and $II$, $\varepsilon_{I,II} (\mathbf r )$ are the (complex) permittivities inside the volume $V$ where the switching action takes place, and $\varepsilon''$ denotes the imaginary part of the permittivity. In most of the cases, the difference $\varepsilon_I (\mathbf{r}) - \varepsilon_{II} (\mathbf{r})$ is non-zero and constant at the sole spatial locations corresponding to the switching material. Hence, the second member of Eq.~1 only depends on its permittivity, defining a \textit{material figure of merit} independent of the specific device shape. Solving the inequality for $\mathrm{min}[T_I,T_{II}]$, the diagram reported in Fig.~1 (b) is obtained. Here exists a forbidden region which is inaccessible by any device built out of a material which has a given $\gamma_{mat}$; in other words, it is the switching material that ultimately dictates the minimum amount of losses introduced by the device into the optical path\footnote{It should be highlighted that the limit expressed by Eq.~1 is reached when the main contribution to the total losses is that originating from the absorption in the switching material. Hence, the reduction of losses such as reflection and scattering, or dissipation in opaque components other than the switching material, is always beneficial.}. A material with a small $\gamma_{mat}$ will necessarily behave as a ``bad'' actuator, while a material with a large $\gamma_{mat}$ can potentially be at the base of a well performing device. A trivial case is that of a transparent material which only changes the refractive index; in this case,  $\gamma_{mat} \rightarrow \infty$, and it is clearly possible to build an ideal phase switch by simply placing the material itself into the optical path. The reverse is more subtle: given a material with $\gamma_{mat} \rightarrow \infty$, a design effort is in general needed to approach the fundamental limit.

To clarify this point, and to check the validity of the general inequality Eq.~1, we analyze the device schematized in the right part of Fig.~1 (b). It simply consists of a waveguide loaded with the switching material; the overlap of the latter with the modal field is $\Gamma$. For a sufficiently weak perturbation\footnote{The weak perturbation approximation can be safely applied to low-contrast structures; however, finite-element simulations showed that it can be applied with a good accuracy also to silicon-on-insulator waveguides, provided that the loading material does not introduce a very large perturbation to the cladding index, or that the field overlap with the loading material is small enough.}, the waveguide effective index is modified by $(n + i \kappa) \cdot \Gamma$ in state $I$, and by $(n + \Delta n + i \kappa + i \Delta \kappa) \cdot \Gamma$ in state $II$ \cite{SynderLove}. Since the length of the loaded section must be $L_{\pi} = \lambda_0 / 2 \Gamma \Delta n$, the transmittances in states $I$ and $II$ are given by the formulas reported in the Figure; notice that in these expressions the dependence on $\Gamma$ cancels out. By extracting a random set of $n$, $\Delta n$, $\kappa$ and $\Delta \kappa$, the blue dots in Fig.~1 (b) are obtained. All these points lie in the allowed region of the graph. A detailed observation reveals that there is a narrow area between the forbidden region and the cloud of blue points which is not filled, and two possible causes for this effect have been identified. First, the waveguide perturbation approximation has been assumed here, and this may result weaker in certain areas of the parameter space. Second, the blue dots follow from the analysis of a specific device geometry, i.e., the loaded waveguide; this choice may result in devices which do not reach the optimality boundary in the small $\gamma_{mat}$ region. A similar behaviour will be also observed in Sect.~3 about amplitude actuators, and a general solution to that will be discussed in detail in Sect.~4.

Here instead we focus on two cases of special interest, which have been referred to in the above. One is that of a material which is nearly transparent in both states; its representative point is labeled (i) on the graph. Specifically, its parameters are $n = 2$, $\Delta n = 1$, $\kappa = 1.5 \times 10^{-3}$, $\Delta \kappa = 0$. This leads to $\gamma_{mat} \simeq 10^5$ and $T_{I} = T_{II} = 0.99$: that is, a nearly-ideal phase delay device with negligible insertion losses.
Consider instead a material characterized by $n = 2$, $\Delta n = 1$, $\kappa = 5 \times 10^{-6}$, $\Delta \kappa = 0.5$. Again, the figure of merit is $\gamma_{mat} \simeq 10^5$, but the insertion loss in state $II$ is large: $T_{II} = 0.04$ [point (ii)]. In essence, when attempting to realize a loaded-waveguide phase actuator device which relies on this material, a very poor performance is obtained. This is because $\Delta \kappa$ is large compared to $\Delta n$, and the loaded waveguide section mostly works as an amplitude switch.

However, even relying on such a material, it is possible to design a phase switch that approaches the limit given by Eq.~1. Consider for instance the device sketched in Fig.~1 (c): it consists of a ring resonator filter loaded by the switching material. While the switching material itself essentially acts as an amplitude switch, the global device implements a $\pi$ phase shift actuator. Indeed, in the transparent state, and for resonant wavelengths, the ring behaves as an all-pass filter which shifts the output phase by $\pi$  (state $I$). In the opaque state, instead, the ring is ``broken'' and no phase shift appears at the output port (state $II$). This is an example which shows the potentiality of the concept of material figure of merit $\gamma_{mat}$ and of Eq.~1: by a proper device design, it is possible to obtain a quasi-ideal phase switch even though at a first glance the material itself is not suited for that purpose. The distance from the zero-insertion loss condition ($IL \simeq 0 \leftrightarrow T \simeq 1$) is here tuned by a device parameter, the coupling efficiency $K$ [see Fig.~1 (c)]; small $K$'s mean less $IL$'s. It should however be noticed that a small $K$, and hence a small $IL$, is accompanied by a narrow bandwidth, a known tradeoff encountered in optical devices based on resonance.

\section{Fundamental limit on the losses of amplitude actuators}
The second problem we address is that of evaluating the performance of an amplitude switch. Its working principle is schematized in Fig.~2 (a): state $I$ is the ``on'' of the device, in the sense that light is not blocked; conversely, state $II$ is the ``off''. An ideal amplitude switch would leave all the radiation pass in state $I$, while completely blocking it in state $II$. Nonidealities are hence described by the insertion loss $IL$ and by the extinction ratio $ER$, usually expressed in dB scale: $IL = -10 \log_{10} T_I$, $ER = -10 \log_{10} T_{II}/T_I$. As for the phase switch, by generalizing the theory reported in Ref.~\cite{TamagnoneNatPhot2014} it can be shown that the following inequality holds:
\begin{equation}
\frac{T_I \left(  \sqrt{T_I/T_{II}} - 1  \right)^2 }
	 {  \left( 1 - T_I  \right)  
		     \left(  T_I/T_{II} - T_I   \right) } \le \gamma_{mat}
\end{equation}
where the material figure of merit $\gamma_{mat}$ only depends on the switching material permittivities in states $I$ and $II$ (see Eq.~1). 

\begin{figure}[tb]
\centering
\includegraphics[width=\linewidth]{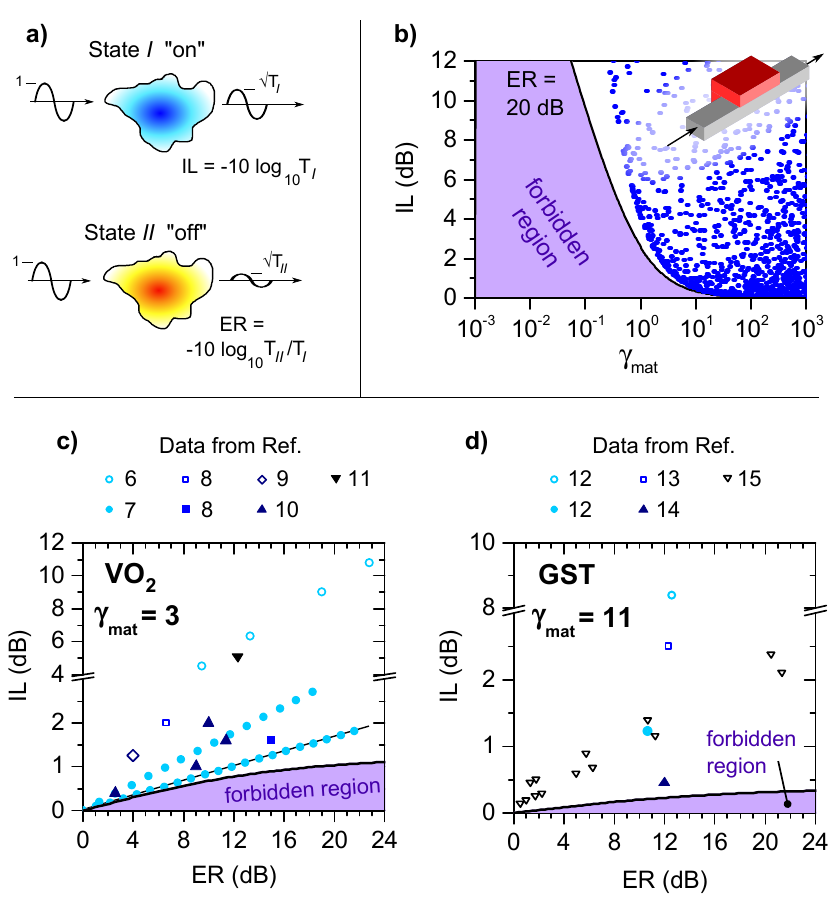}
\caption{Fundamental limit for an amplitude-switching optical element. (a) Schematic of the switching action. (b) Minimum insertion loss as a function of the material figure of merit when an extinction ratio of 20 dB is required. The points represent the loss of a loaded-waveguide intensity switching device, where the refractive index and attenuation coefficient of the material in states $I$ and $II$ is randomly chosen. All the points lie in the allowed region. (c) Validation of the theory based on the analysis of literature data about $\mathrm{VO}_2$. (d) Same as in panel (c), here about GST ($\mathrm{Ge}_2 \mathrm{Sb}_2 \mathrm{Te}_5$). In panels (c) and (d) empty symbols correspond to experimental works, while filled symbols to theoretical ones.}
\end{figure}

Similarly to the result concerning phase actuators, an intensity actuator relying on a material with small $\gamma_{mat}$ will have large insertion losses; conversely, if a material with large $\gamma_{mat}$ is employed, small insertion losses can be obtained. If, for instance, an extinction ratio of 20 dB is required, the limiting curve reported in Fig.~2 (b) applies. Again, the validity of the limit is confirmed by analyzing the performance of the loaded-waveguide device, now designed to act as an intensity switch, in the weak perturbation approximation. Assuming that the complex refractive index of the switching material is $(n + i \kappa)$ in state $I$ and $(n + \Delta n + i \kappa + i \Delta \kappa)$ in state $II$, under this approximation it is straightforward to show that, to achieve an extinction ratio $ER$, the insertion loss is $IL = ER\cdot \kappa/ \Delta \kappa$, independent of the overlap factor $\Gamma$ between the guided mode and the switching material. We extracted a random set of quartets ($n$, $\Delta n$, $\kappa$, $\Delta \kappa$), and represented as a blue dot in Fig.~2 (b) the corresponding pair $(\gamma_{mat},IL)$. All the dots lie in the allowed region, thus confirming the validity of Eq.~2 over a large span of $\gamma_{mat}$.

The support to Eq.~2 reported above however relies on a quite special device geometry and on the weak perturbation approximation; these are also the reasons why the allowed region is not completely filled by the blue points. The discussion about how to get closer to the forbidden region will be systematically addressed in the next Section; here we instead gain further confidence into Eq.~2 by relying on theoretical and experimental results reported in the literature. We have chosen two cases of study, the phase-change materials vanadium dioxide ($\mathrm{VO}_2$) and GST ($\mathrm{Ge}_2 \mathrm{Sb}_2 \mathrm{Te}_5$). These materials attracted much attention in the last years, since the huge contrast which characterizes the optical responses of the two states allows to implement extremely compact devices, with footprints down to submicrometer size. In addition, devices based on these materials are interesting thanks to low energy consumption, to self-holding operation (in the case of GST), and thanks to the integrability of the switching material into existing platforms; most remarkably, into silicon photonics or in connection with surface plasmons. However, most of them suffer from quite large insertion losses, and it naturally arises the question if these losses can be eliminated through a careful design of the devices and technology improvement, or if they are inherent in employing phase change materials. 

In Fig.~2 (c) we plot as dots the insertion losses \textit{vs} the extinction ratios of several $\mathrm{VO}_2$-based devices reported in the literature. Empty marks correspond to experimental works, and filled marks to theoretical ones. All the representative points lie in the allowed region of the graph. It is worth noticing that the results of theoretical works, and especially that of \cite{PoonOE2012}, lie very close to the forbidden region: by relying on vanadium dioxide, no further improvements are possible. Here we employed the value $\gamma_{mat} = 3$, which follows from the complex refractive indices reported in \cite{PoonOE2012}; the values reported in the other articles lead to slightly different $\gamma_{mat}$, but we systematically checked that the corresponding ($IL$, $ER$) values were lying outside the related forbidden region. Similarly, in Fig. 2 (d) we report a set of IL-ER pairs taken from the literature about GST. Here the forbidden region is narrower, in consequence of the fact that GST has a larger $\gamma_{mat}$ with respect to $\mathrm{VO}_2$. Consistently, there are reports in the literature of device performances close to the fundamental limit \cite{PerniceAPL2012}.    

Far from being a complete review of the switching materials employed in integrated optics and nanophotonics, the analyses detailed above show the potentials and limitations of two relevant phase change materials at telecom wavelengths, and provide further confirmation of the validity of Eq.~2.

\section{Resonant versus non-resonant amplitude actuators}
It will now be shown that a switching device whose working principle is non-resonant wave propagation through a region loaded by the absorbing material may be quite far from optimality. Consider, for instance, the family of devices whose representative points are highlighted by a straight line in Fig.~2 (c). These points lie on a straight line since they follow from insertion losses and extinction ratio given per unit length, being the device a plasmonic waveguide loaded by the switching material. Despite the waveguide itself is well optimized (the points are essentially tangent to the curve which delimits the forbidden region), when devices with larger and larger extinction ratio are desired, they turn out to deviate more and more from the fundamental limit. Clearly, this problem is not limited to the VO$_2$-based device of Ref.~\cite{PoonOE2012}; rather, it concerns every switching device based on light propagation through the switching region. While this is not an issue as far as single actuators with low extinction ratios are involved, it may pose a problem in applications where a cascade of actuators or large extinction ratios are needed. 

However, following the limit theory, there are no first-principle limitations to realize a device with insertion losses smaller than those inherent to a component based on wave propagation. Again, as observed above for phase actuators, the key is to base the switch on a resonant element. In Fig.~3 we compare a device based on wave propagation through a simple loaded waveguide with a ring resonator where a section of the loop is replaced by the loaded waveguide. 
\begin{figure}[tb]
\centering
\includegraphics[width=\linewidth]{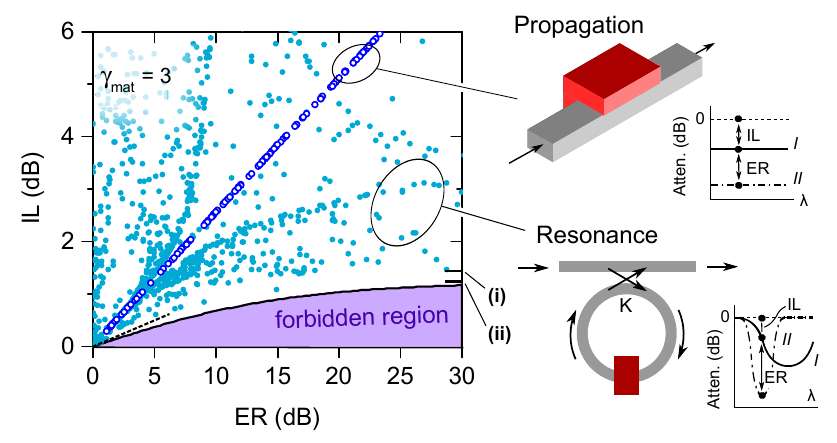}
\caption{Performance of propagation-based and resonance-based amplitude actuators in comparison with the fundamental limit. The resonance-based device can perform better than the propagation-based one, especially in the large $ER$ region. Filled and empty dots are obtained by randomly choosing the key parameters for the two geometries (see text). The tick marked (i) represents the minimum $IL$ achievable at arbitrarily large $ER$ with the ring-based device. The tick marked (ii) represents the minimum $IL$ achievable at arbitrarily large $ER$ for the most general switching device relying on a material with $\gamma_{mat} = 3$.}
\end{figure}
The points describe realistic devices based on a rib Silicon waveguide loaded with VO$_2$, whose geometry is taken from \cite{AtwaterOE2010}. This waveguide is characterized by two complex effective indices, corresponding to the two states of the vanadium oxide: $n_{\mathrm{eff},I} = 2.92$, $n_{\mathrm{eff},II} = 2.68$, $\kappa_{\mathrm{eff},I} = 0.025$, $\kappa_{\mathrm{eff},II} = 0.112$. For a fixed waveguide geometry, and consequently for a given pair of propagation constants $\beta_{I,II} = 2 \pi (n_{\mathrm{eff};I,II} + i \kappa_{\mathrm{eff};I,II}) / \lambda_0$, the only relevant device parameter in the propagation configuration is the length. In the resonant configuration, instead, there are two relevant parameters\footnote{A more refined model would include, for instance, reflection and scattering at the unloaded/loaded waveguide interface, and distributed backscattering. However, since these losses mechanisms can be reduced by a proper engineering, and since the aim is here to analyze the \textit{intrinsic} limits of actuators, these losses are not included in the model.}: the loaded section length $L$ and the intensity coupling coefficient $K$. The total ring length is fixed by imposing the resonance condition either in state $I$ or in state $II$. From the point distribution -- which follows from a random set of the key parameters $L$ and $K$ -- it turns out that, in the large extinction ratio region, the device based on resonance may perform much better than that based on propagation, and that performances very close to the fundamental limit can be obtained. This resonance-mediated approach to the fundamental limit occurs even in the case that the loaded waveguide design by itself is not optimal, which may occur, for instance, due to fabrication constraints. Consider again the data in Fig.~3. Here, the line corresponding to the propagation-based device is not tangent to the forbidden region (the line tangent to the forbidden region, given by $IL = ER \cdot \left(\sqrt{1+1/\gamma_{mat}}-1\right)/2$, is highlighted as a dashed line close to the origin of axes in Fig.~3). Nevertheless, by embedding such a waveguide into a resonant ring, performance much closer to the fundamental limit could be obtained. 

Although for illustrative purposes here we analyzed a VO$_2$-based device, the hint that a resonant device is closer to the fundamental limit than a device based on light propagation will be demonstrated in a general form in the following. To this end, we notice that, in the resonant device, the large extinction ratio regime is reached under the critical coupling condition. Neglecting the bare waveguide transmission losses, one has $ER \rightarrow \infty$ when the coupling between the bus waveguide and the ring is matched with the transmission loss through the loaded section. Consistently with the notation of Fig.~ 2, the material state $II$ has to be chosen as the device ``off'' state; thus, the critical coupling condition is written $K = 1-e^{-2 \beta''_{II} L}$. Given this constraint, the insertion loss at critical coupling is readily obtained in closed form:
\begin{widetext}
\begin{equation}
 IL_{\mathrm{ring}, ER \rightarrow \infty} = -10 \log_{10} \frac{e^{-2\beta''_{I}L} + e^{-2\beta''_{II}L} - 2 e^{-(\beta''_{I} + \beta''_{II}) L} \cos{[(\beta'_{I} - \beta'_{II}) L]}}{1 + e^{-2(\beta''_{I} + \beta''_{II}) L} - 2 e^{-(\beta''_{I} + \beta''_{II}) L} \cos{[(\beta'_{I} - \beta'_{II}) L]}}.
\end{equation}
\end{widetext}
It can be shown (see Supplementary Material) that this expression is minimized when $L \rightarrow 0$, i.e., when $K \rightarrow 0$, and that the limit value is 
\begin{eqnarray}
\lefteqn { IL_{\mathrm{ring}, ER \rightarrow \infty, \mathrm{min}}  = }  \\ \nonumber
&& = -10 \log_{10} \frac{ (\beta''_{I} - \beta''_{II})^2 + (\beta'_{I} - \beta'_{II})^2 }
{ (\beta''_{I} + \beta''_{II})^2 + (\beta'_{I} - \beta'_{II})^2 }.
\end{eqnarray}
The existence of this limit, and the fact that it is finite, is a proof that a critically coupled ring resonator device always outperforms the propagating-wave device, as long as large extinction ratios are considered. The proof given in the Supplementary Material also supports that this conclusion is independent of the specific material under consideration. For the specific case of the VO$_2$-based device analyzed above, this limit is reported as a tick marked (i) in Fig.~3. Consistently, this limit lies below all the points representing the resonant devices at large extinction ratios, while it is above the fundamental limit 
\begin{equation}
IL_{\mathrm{fund}, ER \rightarrow \infty} = -10 \log_{10} \frac{\gamma_{mat}}{1 + \gamma_{mat} }
\end{equation} 
obtained from Eq.~2 and labelled (ii) in Fig.~3. 

It should be noticed that the limit in Eq.~(5) involves the bulk material permittivity (in the case of Fig.~3, VO$_2$), while that in Eq.~(4) involves the propagation constant of the considered waveguide design (in the case of Fig.~3, that of Ref.~\cite{AtwaterOE2010}). However, it is proved in the Supplementary Materials that the limit in Eq.~(4) is always larger than that in Eq.~(5), independently of the specific choice of the switching material and of the waveguide geometry. As it was already noticed in section 2 about phase actuators, the use of resonant components has the drawback that the bandwidth is in general reduced with respect to the case of propagation based devices. Anyway, as far as the optimality with respect to insertion losses are concerned, the results given above together with those given in Sect.~2 support the conclusion that the concept of resonance may play a crucial role in the optimization of optical actuator. Although the discussion in the present article deals with integrated optical waveguides and ring resonators, the generality of the resonance and critical coupling concepts allows to extrapolate the present results also to other photonic platforms such as photonic crystals and metamaterials \cite{TunablePhotonicCrystals, ShalaevLPR2011}.

We conclude this section by noticing that the above analysis does not depend on the choice of the material ``transparent'' state as state $I$ and of the ``opaque'' state as state $II$, or vice-versa. In the deduction of Eq.~2, indeed, this assumption has not been made, and the designer is free to choose the switching material ``opaque'' state for the device ``on'' state (i.e., the device state which does not block the light flow), or the opposite. This fact may be exploited in view of energy saving. Suppose that the need is to design a device intended for normally-on operation, and that the switching material has one of the two states which is power-hungry. The device can be engineered to use the power-hungry material state for the device ``off'' state, hence reducing the overall energy consumption. While this conclusion is general and holds for arbitrary device geometry, it can be read out directly in the framework of the critically coupled ring resonator by noticing that Eqs~3-4 are invariant for the exchange $I \leftrightarrow II$.

\section{Comparison of different materials employed in actuators}
The power of the limits expressed by the inequalities given in Eqs.~1-2 is that it is sufficient to know the figure of merit $\gamma_{mat}$ of the (bulk) switching material to have significant insights into the potentiality of a new material, prior to directly designing specific devices. Furthermore, the limits may be of help as far as an optimization is concerned, when the decision whether to proceed with further optimization steps has to be taken. It is clear that the inequalities given above and the material figure of merit only provide information on a single metric on the device performance, while other issues like bandwidth, footprint, switching energy, state retention, switching time etc.\ are not grasped by $\gamma_{mat}$. Nevertheless, the knowledge of $\gamma_{mat}$ could be of help, for instance, in choosing the material which is best suited for operation in a certain wavelength range. Indeed, $\gamma_{mat}$ only depends on the permittivities, which are often known from optical reflectometry or ellipsometry, from first-principle structural calculations, or from other models. 

In Fig.~4 we propose this spectral comparison, regarding two phase change materials ($\mathrm{Ge}_2 \mathrm{Sb}_2 \mathrm{Te}_5$, referred to as GST, and $\mathrm{VO}_2$), a transparent conductive oxide (Indium Tin Oxide, ITO), and a semiconductor (Silicon). 
\begin{figure}[htb!]
\centering
\includegraphics[width=\linewidth]{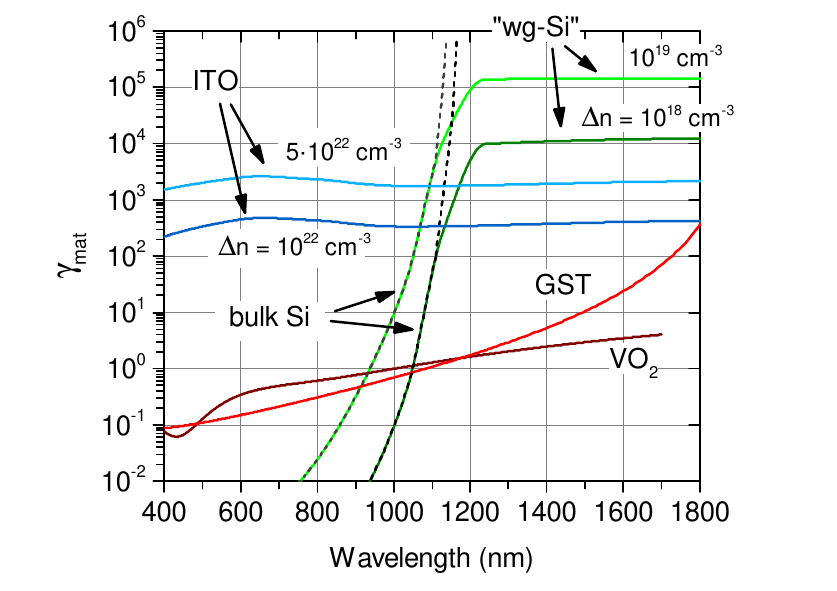}
\caption{Spectral dependence of the figure of merit for four materials employed in nanophotonics, whose working principle is different. Dielectric modulation in $\mathrm{VO}_2$ and GST ($\mathrm{Ge}_2 \mathrm{Sb}_2 \mathrm{Te}_5$) is due to a phase transition, while in ITO and Silicon the plasma effect due to free charge is considered. In Silicon, for wavelengths longer than the bandgap, extrinsic losses due to waveguide scattering are included.}
\end{figure}
In the phase change materials, the permittivity change is induced by a structural transition -- amorphous-crystalline in the case of GST, and from a monoclinic to a rutile structure in the case of $\mathrm{VO}_2$. The dielectric functions are retrieved from \cite{TazawaJJAP2007, OravaAPL2008}. In the case of ITO, the plasma dispersion effect modeled by a Drude contribution to the permittivity is responsible for the modulation effect. Here, typical parameters for the dielectric response are taken from \cite{Volker, AtwaterNanoLett2010, DobrowolskiAO1983, MichelottiOL2009, Jung} and correspond to a mobility of $15\ \mathrm{cm}^2/\mathrm{Vs}$. As opposed to the phase change materials, whose response is intrinsic to their structure (a change in certain optical matrix elements for GST \cite{Parrinello}, and a semiconductor-insulator Mott transition for $\mathrm{VO}_2$ \cite{Basov}), the plasma effect in ITO can be tuned through the electron population injected or accumulated in the active region. It turns out that the material figure of merit significantly depends on that parameter, gaining more than one order of magnitude over a wide spectral range for an order-of-magnitude change in the electron density. 

The plasma dispersion effect is also at the origin of the response of Silicon \cite{SorefIEEE1987}, and is here quantified assuming a mobility of $1500\ \mathrm{cm}^2/\mathrm{Vs}$, and an injected electron density of $10^{18}$ or $10^{19}\ \mathrm{cm}^{-3}$. By introducing also the effect of holes the figure of merit is increased by a factor $\sim 2$. As opposed to the other materials, which have a flat response in a wide spectral range, Silicon strongly feels the effect of a bandgap. If the bulk Si permittivity is employed, in the ``undoped'' state the material is well transparent, implying values of $\gamma_{mat}$ larger than $10^6$ above the $1.2\ \mu$m wavelength. However, when Silicon is employed for optical waveguides, extrinsic losses due to roughness scattering and surface state absorption always occur. These losses, despite being extrinsic to the bulk material, and rather connected to the device itself, can however be accounted for in the material figure of merit, defining a $\gamma_{mat}$ for an effective ``waveguide-Silicon'' material. Assuming a loss of $1\ \mathrm{dB/cm}$ \cite{MorichettiPRL2010}, values of $\gamma_{mat} = 10^{4} - 10^5$, flat in the whole near-infrared spectral range, are obtained. If instead a low-loss $0.1\ \mathrm{dB/cm}$ Si waveguide is considered \cite{BibermanOL2012}, the material figure of merit increases by an order of magnitude. As for ITO, also in Silicon the figure of merit depends significantly on the injected charge density. Hence, provided that the mobility is not reduced when a large charge density is involved, it is convenient to work in this regime. This is a consequence of the balance between the real and imaginary part of the permittivity given by the Drude model, and applies to every material whose switching action relies on this mechanism.

\section{Conclusions and perspectives}
In conclusion, we derived fundamental limits on the losses of arbitrarily shaped two-port amplitude and phase optical actuators. Finding their roots into a simple manipulation of Maxwell equations for linear and reciprocal dielectrics, the validity of these limits extends to a wealth of linear switching devices, and in particular to integrated optics devices regardless of the specific geometric configuration. The key role is played by the switching material, whose effectiveness is quantified by a material figure of merit simply defined in terms of the permittivities. While the introduced figure of merit does not give insights into certain metrics like switching time, footprint, state retention, switching energy etc., it sets clear limits on the optical performances of any device which relies on a given material. Further, we observed a peculiar connection between the ability to reach the fundamental limit and the presence of resonance and of critical coupling in the operation principle of the device. We believe that the present theory provides an important metric tool which will direct researchears towards highly performing optical devices and materials. 

\section*{Acknowledgements}
The research leading to these results has received funding from the EU Seventh Framework Programme (FP7/2007-2013) under grant agreement number 323734 ``Breaking the Barrier on Optical Integration'' (BBOI). Fruitful discussions with Daniele Melati and Marco Morandotti are also gratefully acknowledged.

\section*{Journal reference}
This article is published in \textit{Laser and Photonics Reviews} \textbf{9}, No.~6, 666 (2015). DOI: 10.1002/lpor.201500101

\end{document}


\title{Supplementary Material for the article \textit{Fundamental limits on the losses of phase and amplitude optical actuators} }

\author{Simone Zanotto}
\email{simone.zanotto@nano.cnr.it}
\affiliation{Dipartimento di Elettronica, Informazione e Bioingengeria, Politecnico di Milano, P.za Leonardo da Vinci 32, 20133 Milano, Italy.}
\altaffiliation{Present address: Istituto Nanoscienze - CNR, and Laboratorio NEST, Piazza San Silvestro 12, 56127 Pisa, Italy}

\author{Francesco Morichetti}
\affiliation{Dipartimento di Elettronica, Informazione e Bioingengeria, Politecnico di Milano, P.za Leonardo da Vinci 32, 20133 Milano, Italy}

\author{Andrea Melloni}
\affiliation{Dipartimento di Elettronica, Informazione e Bioingengeria, Politecnico di Milano, P.za Leonardo da Vinci 32, 20133 Milano, Italy}

\date{\today}

\begin{abstract}
This Supplementary Material provides the formal theoretical background underlying the fundamental inequalities discussed in the main article.
\end{abstract}

\maketitle

\section{General scattering inequality for a multiport, dielectrically-driven optical actuator}

Consider a generic two-state optical system like that represented in Fig.~S1.
\begin{figure}[tb]
\centering
\includegraphics[width=\linewidth]{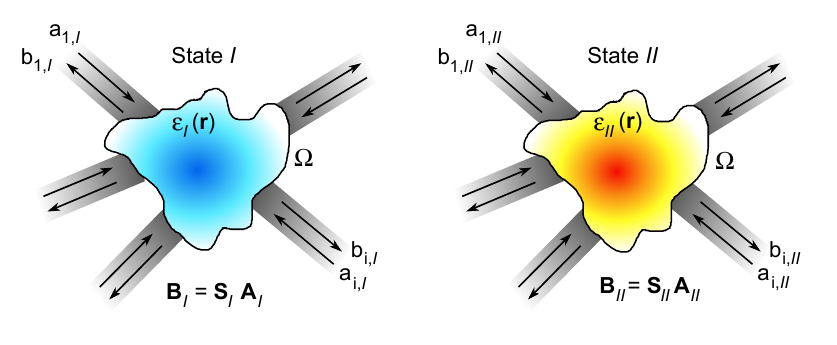}
\caption{Schematic of a two-state switching optical element. In the linear optical response regime it is fully characterized by the scattering matrix $\mathbf S$, which connects the input amplitudes with the output amplitudes. The switching action is eventually ascribed to a modulation of the dielectric constant spatial distribution $\varepsilon (\mathbf{r})$.}
\end{figure}
An ``active region'' is connected to the exterior via a number of waveguides, where ingoing and outgoing fields are quantified by the complex amplitudes $\mathbf A = (a_1 \ldots a_i \ldots)$ and $(\mathbf B = b_1 \ldots b_i \ldots)$, respectively. The harmonic time dependence $e^{i \omega t}$ is assumed; disregarding non-linear phenomena, the system's response is fully described by the scattering matrix $\mathbf S$ that connects $\mathbf A$ with $\mathbf B$. Actually, the amplitudes $a_i$ does not necessarily correspond to spatially separated waveguide channels, the only requirement is modal orthogonality between the (discrete) scattering channels. Hence, the present theory applies to more general systems, like periodic arrays (where plane waves corresponding to open diffraction channels are involved), systems with spherical or cylindrical symmetry (where Bessel spatial harmonics are involved), or mode-division multiplexing devices.

It will be now shown that the changes in the system as seen from the exterior, i.e., the device switching performance described by a change in the $S$-parameters from state $I$ to state $II$, are fundamentally connected to the values that the permittivity of the materials contained in the ``active region'' assume in the two states. 

Starting point is the expansion into normal modes of the transverse electromagnetic fields corresponding to the scattering channels. Referring again to Fig.~S1, we define a surface $\Omega$ which encloses the device. The position of $\Omega$ is chosen to be sufficiently far away from the ``active region'': in this way, on $\Omega$ the electromagnetic fields are only those corresponding to the guided modes of the waveguides, which are supposed to orthogonally enter $\Omega$. The electromagnetic field of each mode $\mu$ at $\Omega$ is hence fully described by the tangential components of the fields\cite{WhatIs}:

\begin{eqnarray}
 \mathbf E_{T,\mu} (x,y,z) & = & (a_{\mu} e^{-i \beta_{\mu} z} + b_{\mu} e^{i \beta_{\mu} z})\ \mathbf e_{T,\mu}(x,y) \nonumber \\
 \mathbf H_{T,\mu} (x,y,z) & = & (a_{\mu} e^{-i \beta_{\mu} z} - b_{\mu} e^{i \beta_{\mu} z})\ \mathbf h_{T,\mu}(x,y)
 \label{fields_expansion}
\end{eqnarray}
where $z$ is a local coordinate pointing into the $\Omega$ surface, and $\mathbf e_{T,\mu}(x,y)$, $\mathbf h_{T,\mu}(x,y)$ are the modal field profiles. Here, $\mu$ is an index which may refer to the modes of different waveguides, or of the same waveguide; in both cases, the orthogonality relation  
\begin{equation}
\int_{\Omega} \left( \mathbf e_{\mu} \times \mathbf h^*_{\nu} \right) \cdot \ud \mathbf n= 2 \delta_{\mu, \nu}
\end{equation}
holds. With $\ud \mathbf n$ we identify the surface element with normal pointing into $\Omega$. We also assume that the waveguides have real and isotropic permeability and permittivity, and hence that the modal transverse field profiles can be chosen as real\cite{Tamir}. 

Consider the sourceless Maxwell equations for the situation \textit{I}:
\begin{eqnarray}
\nabla \times \mathbf E_I & = & -i \omega \mu_0    \mathbf H_I  \label{maxw1} \\ 
\nabla \times \mathbf H_I & = & i \omega \eps_I \mathbf E_I \label{maxw2}
\end{eqnarray}
where $\mathbf E_I$ and $\mathbf H_I$ are the fields corresponding to the excitation vector $\mathbf A_I$, and to the internal device configuration described by the permittivity $\eps_I$. We also suppose that all the materials do not have a magnetic response. 

By dot-multiplying on the left Eq.~(\ref{maxw1}) with $\mathbf H_{II}$ and Eq.~(\ref{maxw2}) with $\mathbf E_{II}$, and summing, we get:
\begin{equation}
\mathbf H_{II} \cdot (\nabla \times \mathbf E_I) + \mathbf E_{II} \cdot (\nabla \times \mathbf H_I) = i \omega (\mathbf E_{II}  \eps_I \mathbf E_I - \mathbf H_{II} \mu_0 \mathbf H_I)
\label{intermedio1}
\end{equation}
Rewriting Eqs.~(\ref{maxw1}-\ref{maxw2}) for ``\textit{II}'', and multiplying for the fields ``\textit{I}'', an expression similar to Eq.~(\ref{intermedio1}) is obtained. Summing up these intermediate results, the following lemma is obtained:
\begin{equation}
\nabla \cdot \left( \mathbf E_I \times \mathbf H_{II} - \mathbf E_{II} \times \mathbf H_I \right) = i \omega \left( \mathbf E_{II} ( \eps_I - \eps_{II} ) \mathbf E_I \right)
\end{equation}
This expression has now to be integrated over the volume $V$ enclosed in $\Omega$. The integral of the left-hand side is connected with the scattered amplitudes via the equivalence
\begin{eqnarray}
\int_V \nabla \cdot \left( \mathbf E_I \times \mathbf H_{II} - \mathbf E_{II} \times \mathbf H_I \right) \ud \mathbf r = \nonumber \\ 
= - \int_{\Omega}   \left( \mathbf E_I \times \mathbf H_{II} - \mathbf E_{II} \times \mathbf H_I \right) \cdot \ud \mathbf n = \nonumber \\ 
=  - 4 \sum_{\mu} \left( b_{\mu, I} a_{\mu, II} - a_{\mu, I} b_{\mu, II} \right) = \nonumber \\
- 4 \mathbf A_{II}^T ( \mathbf S_I - \mathbf S_{II}^T ) \mathbf A_I
\end{eqnarray}
hence resulting in
\begin{equation}
   4 \mathbf A_{II}^T ( \mathbf S_I - \mathbf S_{II}^T ) \mathbf A_I = -i \omega \int_{V} \mathbf E_{II} ( \eps_I - \eps_{II} ) \mathbf E_I\ \ud \mathbf r \label{uno}.
\end{equation}

This result connects the input/output amplitudes for a given excitation configuration in the system states \textit{I} and \textit{II} with the local field distribution inside the interaction region and the material properties described by the permittivity. Similarly, we need to link the device losses as seen from the scattering channels with the power dissipation caused by the presence of a lossy dielectric. From the definitions of the scattering amplitudes $\mathbf A$ and $\mathbf B$, and the dipolar dissipation formula \cite{Jackson}, one obtains
\begin{equation}
\mathbf A_I^H (\mathbf I - \mathbf S_I^H \mathbf S_I ) \mathbf A_I = - \frac{i \omega}{4} \int_V \mathbf E_I^* (\eps_I - \eps^*_I) \mathbf E_I  \label{due}
\end{equation}
and similarly for state $II$, where \textit{H} stands for the Hermitian conjugate and $\mathbf I$ is the identity matrix.

Equations \ref{uno} and \ref{due} can be introduced in a chain of inequalities proved in\cite{TamagnoneNatPhot2014}, giving rise to a device and a material figure of merit linked by an inequality:
\begin{equation}
\gamma_{dev} \equiv \frac{|\mathbf A_{II}^T ( \mathbf S_I - \mathbf S_{II}^T ) \mathbf A_I|^2}
             {\left[ A_I^H (\mathbf I - \mathbf S_I^H \mathbf S_I ) \mathbf A_I \right] \left[ A_{II}^H (\mathbf I - \mathbf S_{II}^H \mathbf S_{II} ) \mathbf A_{II} \right]} \label{gammadev}
\end{equation}

\begin{equation}
\gamma_{mat} \equiv \smash{\displaystyle\max_{\mathbf r \in V} \frac{|\varepsilon_I (\mathbf{r}) - \varepsilon_{II} (\mathbf{r})|^2}{4\  \varepsilon_I'' (\mathbf{r}) \ \varepsilon_{II}'' (\mathbf{r})} } \label {gammamat} 
\end{equation}

\begin{equation}
\gamma_{dev} \le \gamma_{mat}. \label{fundamental}
\end{equation}
In Eq.~(\ref{gammamat}) the maximum is taken over the whole active region, and $\eps'' = (\eps - \eps^*)/2i$. The equations above generalize those reported in \cite{TamagnoneNatPhot2014} to an arbitrarily-shaped, dielectrically driven multiport switching device.

\section{Fundamental limit for a phase optical actuator}

Starting from Eqs.~(\ref{gammadev}-\ref{fundamental}), and setting $\mathbf A_I = (1, 0)^T$, $\mathbf A_{II} = (0, 1)^T$ one has
\begin{eqnarray}
\gamma_{dev} = \frac{\left| |t_I|e^{i \phi_I} - |t_{II}| e^{i \phi_{II}} \right|^2}{\left( 1 - |r_I|^2 - |t_{I}|^2 \right)
\left( 1 - |r_{II}|^2 - |t_{II}|^2 \right)} \\
= \frac{\left| |t_I| + |t_{II}| \right|^2}{\left( 1 - |r_I|^2 - |t_{I}|^2 \right)
\left( 1 - |r_{II}|^2 - |t_{II}|^2 \right)}
\end{eqnarray}
if a $\pi$ phase switch, operating in transmission, is required. Here, $t_{I,II}$ and $r_{I,II}$ are the complex amplitude transmission and reflection coefficients, and $\phi_{I,II}$ the transmission phases. Introducing the intensity transmittance $T_{I,II} = |t_{I,II}|^2$, the following inequalities hold:
\begin{eqnarray}
\gamma_{dev} \ge \frac{\left| \sqrt{T_I} + \sqrt{T_{II}} \right|^2}{\left( 1 - T_I \right) \left( 1 - T_{II} \right)} \\
\ge \frac{\left| \sqrt{T_I} + \sqrt{T_{II}} \right|^2}{(1-\mathrm{min}[T_I,T_{II}])^2} \\
\ge \frac{\left| 2 \sqrt{ \mathrm{min}[T_I,T_{II}] } \right|^2}{(1-\mathrm{min}[T_I,T_{II}])^2} \\
= \frac{4\ \mathrm{min}[T_I,T_{II}]}{(1-\mathrm{min}[T_I,T_{II}])^2}
\end{eqnarray}
from which, in conjunction with (\ref{fundamental}), Eq.~1 of the main article is obtained.

\section{Fundamental limit for an amplitude actuator}

The fundamental limit for an amplitude actuator, Eq.~2 of the main article, is obtained from Eqs.~(\ref{gammadev}-\ref{fundamental}). This generalizes the results reported in \cite{TamagnoneNatPhot2014} to an arbitrarily-shaped, dielectrically driven optical amplitude actuator.

\section{Limits for a critically-coupled ring-resonator amplitude actuator}
In the following we will provide a numerical proof of the statements reported in the discussion about Equations 3, 4 and 5 in the main text.

To the sake of clarity, we recall here the relevant quantities:
\begin{eqnarray}
  \lefteqn {IL_{\mathrm{ring}} =} \\ \nonumber
 &&-10 \log_{10} \frac{e^{-2\beta''_{I}L} + e^{-2\beta''_{II}L} - 2 e^{-(\beta''_{I} + \beta''_{II}) L} \cos{((\beta'_{I} - \beta'_{II}) L)}}{1 + e^{-2(\beta''_{I} + \beta''_{II}) L} - 2 e^{-(\beta''_{I} + \beta''_{II}) L} \cos{((\beta'_{I} - \beta'_{II}) L)}}. \label{ring1}
\end{eqnarray}
\begin{equation}
IL_{\mathrm{ring}, \mathrm{min}}  = 
-10 \log_{10} \frac{ (\beta''_{I} - \beta''_{II})^2 + (\beta'_{I} - \beta'_{II})^2 }
{ (\beta''_{I} + \beta''_{II})^2 + (\beta'_{I} - \beta'_{II})^2 }. \label{ring2}
\end{equation}
\begin{equation}
IL_{\mathrm{fund} } = -10 \log_{10} \frac{\gamma_{mat}}{1 + \gamma_{mat} }
\label{ring2}
\end{equation} 
where the notation $ER \rightarrow \infty$ reported in the main text is not repeated for short.

\begin{figure*}[t]
\centering
\includegraphics[width = 11 cm]{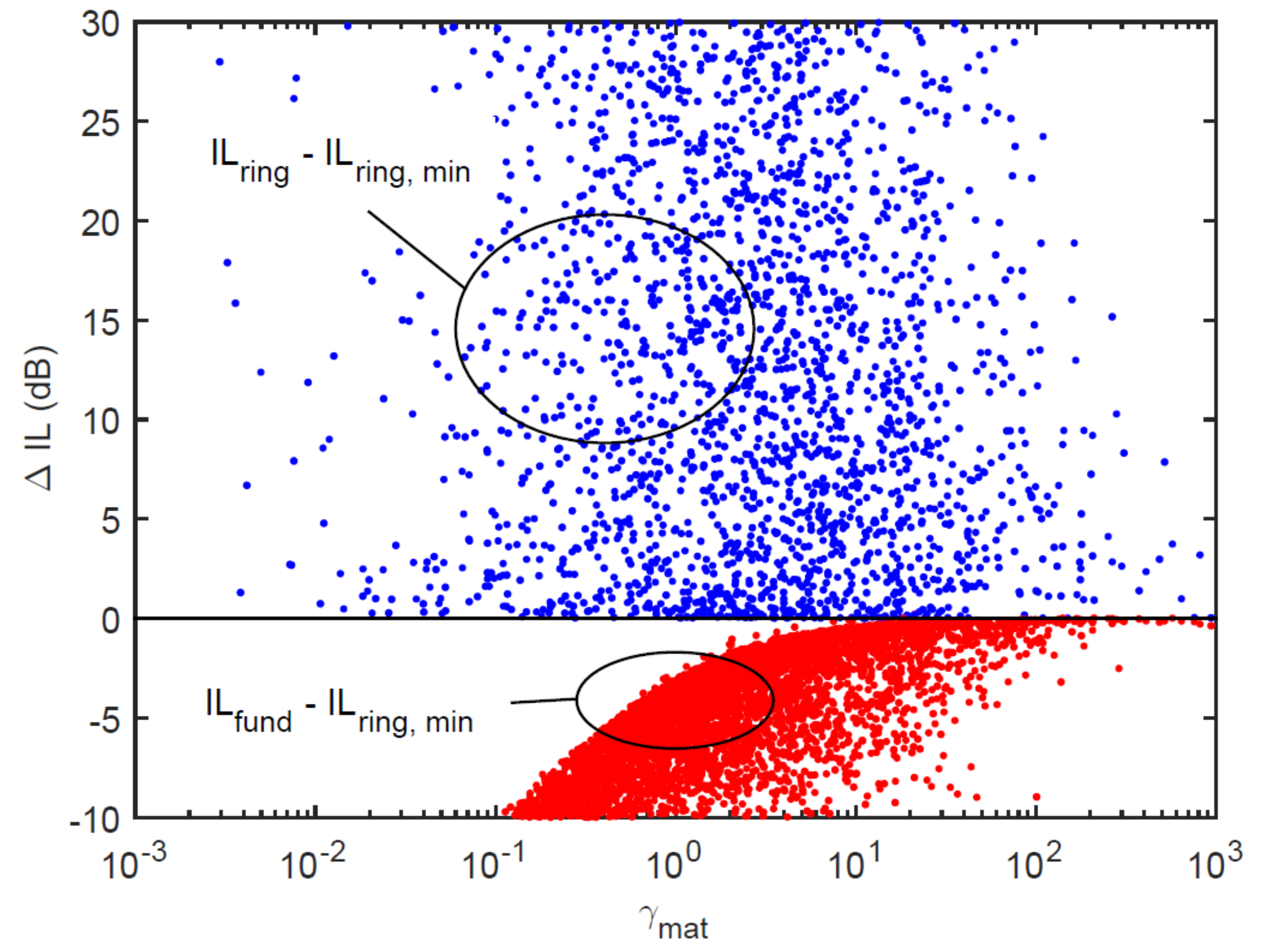}
\caption{Numerical proof of Eq.~\ref{ring_ineq} concerning the limits on critically coupled ring resonator amplitude actuators.}
\end{figure*}

We will now prove that 
\begin{equation}
IL_{\mathrm{ring}} > IL_{\mathrm{ring}, \mathrm{min}} > IL_{\mathrm{fund} } 
\label{ring_ineq}
\end{equation}
for a wide range of values of $\beta'_{I}$, $\beta'_{II}$, $\beta''_{I}$, $\beta''_{II}$, and $L$. To this extent, we first restate the problem in terms of a set of normalized variables. In the weak perturbation approximation for a loaded waveguide, one has $\beta_{I,II} L = 2 \pi \left( n_{\mathrm{eff}} + \Gamma (n_{I,II} + i \kappa_{I,II}) \right) L/\lambda$, where $n_{\mathrm{eff}}$ is the unloaded waveguide effective index, $\Gamma$ is the overlap factor between the modal field and the switching material, and $n_{I,II} + i \kappa_{I,II}$ is the complex refractive index of the switching material. If the propagation constants are rewritten as follows
\[
\beta_{I,II} L = 2 \pi \left( 1 + \tilde{n}_{I,II} + i \tilde{\kappa}_{I,II}) \right) \tilde{L} ,
\]
with $\tilde{n}_{I,II} = \frac{\Gamma}{n_{eff}} n_{I,II} $, $\tilde{\kappa}_{I,II} = \frac{\Gamma}{n_{eff}} \kappa_{I,II} $, $ \tilde{L} = \frac{L n_{eff}}{\lambda}$, the material figure of merit results
\[
\gamma_{mat} = \frac{\left| (\tilde{n}_I + i \tilde{\kappa}_I)^2 - (\tilde{n}_{II} + i \tilde{\kappa}_{II})^2 \right|^2}
{4\  \mathrm{Im} \left[ (\tilde{n}_I + i \tilde{\kappa}_I)^2 \right] \mathrm{Im} \left[ (\tilde{n}_{II} + i \tilde{\kappa}_{II})^2 \right]}.
\]
In essence, the independent variables are the five real numbers $\tilde{n}_{I,II}$, $\tilde{\kappa}_{I,II}$, and $\tilde{L}$.  By allowing these variables to randomly and independently sweep the interval $[0.01,1]$, we obtained a set of values of $IL_{\mathrm{ring}}$,  $IL_{\mathrm{ring}, \mathrm{min}}$, and $ IL_{\mathrm{fund} }$. By plotting the values $IL_{\mathrm{ring}} - IL_{\mathrm{ring}, \mathrm{min}}$ and $ IL_{\mathrm{fund} } - IL_{\mathrm{ring}, \mathrm{min}}$ versus $\gamma_{mat}$ (see Fig.~S2), we get strong numerical evidence for Eq.~\ref{ring_ineq}.